\documentclass[aps,prl,twocolumn,showpacs,superscriptaddress, preprintnumbers]{revtex4-1}
\usepackage{latexsym}
\usepackage{amssymb}
\usepackage{graphicx}
\usepackage{amsmath}
\usepackage{ulem}
\usepackage{bm}
\usepackage[colorlinks,
          linkcolor=black,
            citecolor=black,
            urlcolor=blue
           ]{hyperref}
\usepackage{verbatim}
\usepackage{mathrsfs}
\usepackage{extarrows}
\usepackage{comment}
\usepackage{mathtools,slashed}
\usepackage{soul}
\usepackage[toc,page]{appendix}
\usepackage[vcentermath]{youngtab}
\usepackage{multirow}
\usepackage{atbegshi,picture}
\usepackage{lipsum}
\usepackage{bbm}

\begin{document}

\title{Gappability Index for Quantum Many-Body Systems}

\author{Yuan Yao}
\email{yuan.yao@riken.jp}
\affiliation{Condensed Matter Theory Laboratory, RIKEN CPR, Wako, Saitama 351-0198, Japan}

\author{Masaki Oshikawa}
\affiliation{Institute for Solid State Physics, The University of Tokyo. Kashiwa, Chiba 277-8581, Japan}
\affiliation{Kavli Institute for the Physics and Mathematics of the Universe (WPI),
The University of Tokyo, Kashiwa, Chiba 277-8583, Japan}
\affiliation{Trans-scale Quantum Science Institute, University of Tokyo, Bunkyo-ku, Tokyo 113-0033, Japan}

\author{Akira Furusaki}
\affiliation{Condensed Matter Theory Laboratory, RIKEN CPR, Wako, Saitama 351-0198, Japan}
\affiliation{Quantum Matter Theory Research Team, RIKEN CEMS, Wako, Saitama 351-0198, Japan}

\date{\today}

\begin{abstract}
We propose an index $\mathcal{I}_G$ which characterizes the degree of gappability, namely the difficulty to induce a unique ground state with a nonvanishing excitation gap,
in the presence of a symmetry $G$.
$\mathcal{I}_G$ represents the dimension of the subspace of ambient uniquely-gapped theories in the entire $G$-invariant ``theory space''.
The celebrated Lieb-Schultz-Mattis theorem corresponds, in our formulation, to the case $\mathcal{I}_G=0$ (completely ingappable)
for the symmetry $G$ including the lattice translation symmetry.
We illustrate the usefulness of the index by discussing the phase diagram of spin-1/2 antiferromagnets in various dimensions, which do not necessarily have the translation symmetry.
\end{abstract}

\maketitle

\textit{Introduction.---}
Quantum critical phenomena have been a central subject of physics.
Generic quantum many-body systems are expected to have spontaneously symmetry-breaking (SSB) ground states or a unique ground state below a non-vanishing gap; a parameter fine-tuning would be required to
reach a quantum critical point.
Interestingly, however, in quantum many-body systems, the concept of ``ingappability'' has been developed in the context of
Lieb-Schultz-Mattis (LSM) theorem~\cite{Lieb:1961aa} and its generalizations~\cite{Affleck:1986aa,OYA1997,Oshikawa:2000aa, Hastings:2004ab}:
translationally invariant systems under certain symmetry conditions must have either a gapless spectrum above ground state(s) or a ground-state degeneracy.
In such systems, gapped phases with a unique ground state (which we shall call ``uniquely gapped'' phases for short) are excluded from phase diagrams,
while gapless critical phases acquire enhanced stability.

LSM-type theorems only tell whether the system can be uniquely gapped under spatial symmetry such as the translation symmetry.
In ``gappable'' systems not constrained by the LSM-type theorems,
gapless phases are expected to {be less stable} and may even disappear from phase diagrams.
Nevertheless, a large number of critical phases have been observed numerically and experimentally~\cite{Alcaraz:1989aa,Fuhringer:2008aa,Matsumoto-YbAlB4_Science2011}.
Their existence suggests a refined notion of the (in)gappability.

\begin{figure}[t]
\centering
\includegraphics[width=8.8cm,pagebox=cropbox,clip]{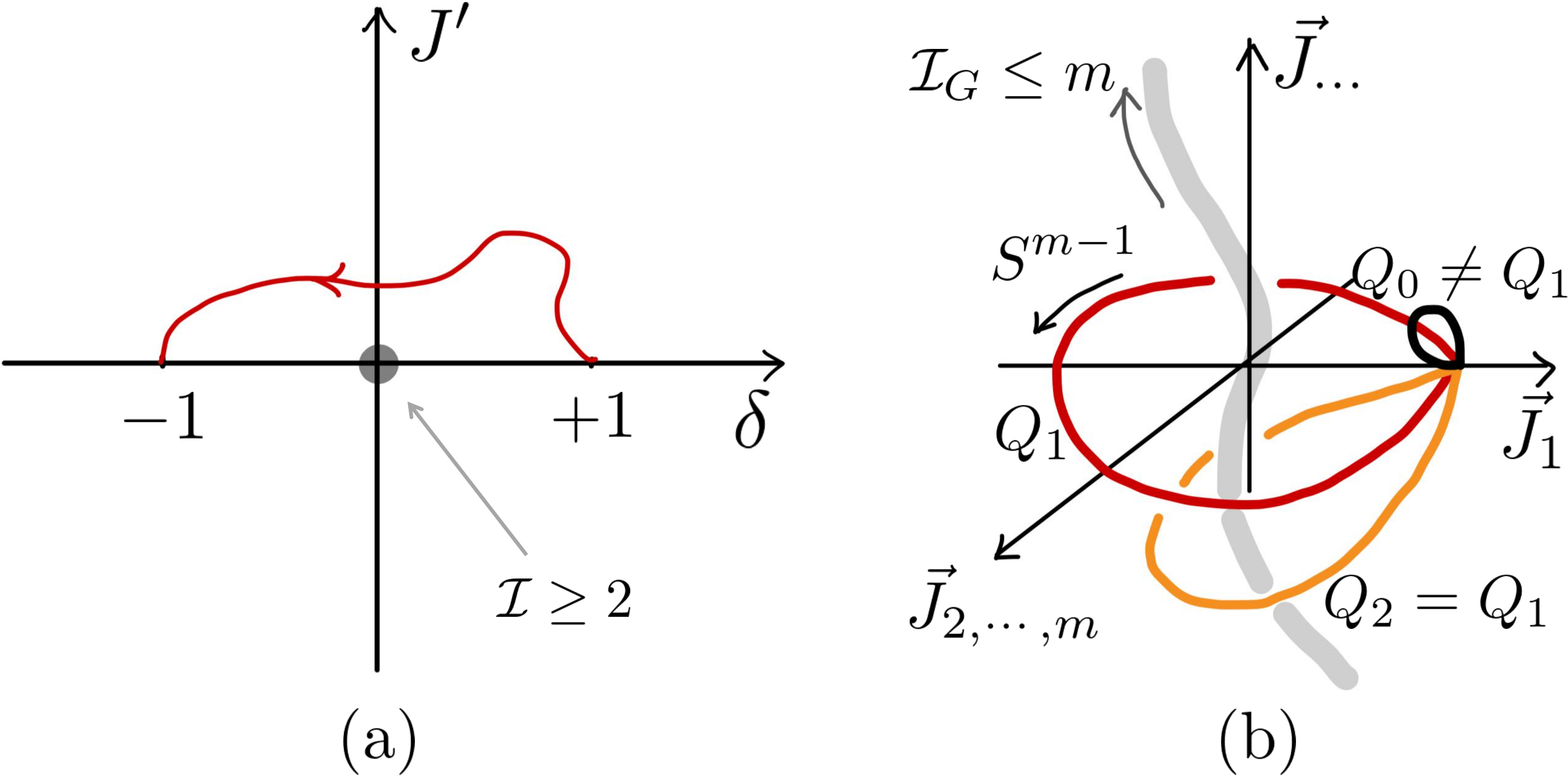}
\caption{(a) $\mathcal{I}$'s of NUG Hamiltonians (denoted by dots) in a three-dimensional parameter space. Here the curve, surface and solid cube represent the NUG phase to which a NUG Hamiltonian belongs while blank {regions} are uniquely-gapped phases. (b) A path of uniquely-gapped Hamiltonians exists once the dimension extension of the NUG phase ceases. (c)~Two $S^{m-1}$'s with distinct topological assignments are not deformable to each other and an extending NUG phase with $\mathcal{I}_G\leq m$ obstructs the contractibility of $S^{m-1}$ if $Q_{1,2}\neq Q_0$.}
\label{method}
\end{figure}

In this Letter,
we propose an integer index $\mathcal{I}_G$ to characterize the degree of gappability of a non-uniquely-gapped (NUG) Hamiltonian respecting a symmetry $G$, where NUG means having gapless low-energy excitations, {SSB or fractionalizations}.
For each NUG Hamiltonian with a symmetry $G$ denoted by a point in a parameter space $P$,
we define its $\mathcal{I}^P_G$ as the codimension of contiguous NUG phases illustrated in FIG.~\ref{method}(a), {where} $\mathcal{I}^P_G$
can be understood as the number of gapping directions~\footnote{Here and after, ``gapping'' means gapping the NUG Hamiltonian to be uniquely-gapped}.
However,
$\mathcal{I}^P_G$ depends on the chosen parameter space $P$ {with a particular set of coupling constants},
so it is meaningful to introduce \textit{the} complete, infinite-dimensional parameter space whose coordinate axes exhaust \textit{all} $G$-symmetric coupling constants \footnote{Another situation, where considering such a complete parameter space is necessary and natural, is the renormalization-group transformations during which infinitely-many interactions are generated in general.}, 
and we denote $\mathcal{I}^P_G$ on it as $\mathcal{I}_G$.
Thus experimental phase diagrams {with a finite number of parameters} are its various sections with $\mathcal{I}^P_G \leq\mathcal{I}_G$~\footnote{As more and more interaction couplings are introduced into the parameter space, the gapping directions of the certain NUG Hamiltonian cannot decrease.}
and a greater $\mathcal{I}_G$ implies being more gappable.
The original LSM-type ingappability corresponds to $\mathcal{I}_G=0$ representing the absence of uniquely-gapped phases.
$\mathcal{I}_G$ provides more refined constraints on quantum phase diagrams,
and give finer measures of critical-phase/point stability than the LSM-type theorem that only indicates $\mathcal{I}_G=0$ or not.


Our notion of (in)gappabilities indicated by $\mathcal{I}_G$ is related to the codimension of topologically protected gapless defects/boundaries in gapped phases of free fermions~\cite{Teo:2010aa,Ryu2010}
and that of gapless points in the phase diagram of field theories~\cite{Cordova:2020aa,Cordova:2020ab,Kapustin:2020aa,Kapustin:2020ab,Hsin:2020aa}.
Here, {as concrete examples,}
we consider quantum spin systems on $d$-dimensional lattices that may not admit a simple description in terms of non-interacting particles or a known field theory.
We claim the following theorem as our main result:

\textit{Ingappability of spin-1/2 antiferromagnets: There exists a quantum phase of Hamiltonians with $\mathcal{I}_G\leq d-k$ in spin-1/2 antiferromagnetic systems on $d$-dimensional cubic lattices and $G=G_\text{onsite}\times(\mathbb{Z})^k$ where $G_\text{onsite}$ is an onsite symmetry that is one of i) SO$(3)$ spin-rotation symmetry, ii) {dihedral symmetry of $\pi$-spin-rotations} $\mathbb{Z}_2\times\mathbb{Z}_2$, or iii) a time-reversal symmetry $\mathbb{Z}_2^{\mathcal{T}}$. 
$(\mathbb{Z})^k$ denotes the translational symmetry along $k\leq d$ lattice direction(s).}

Since the case of $k=d$ is reduced to the LSM-type theorems~\cite{Chen-Gu-Wen_classification2010,Fuji:2016aa,Watanabe:2015aa,Cheng:2016aa,Po:2017aa,Ogata:2018aa,Ogata:2020aa,Else:2020aa,Yao:2021aa}, 
we focus on $k<d$.
To prove the statement,
we first consider the extreme case $k=0$ or $G=G_\text{onsite}$ without lattice {translation symmetry} required, and take the translations into consideration later.
In the following parts,
we obtain the above statement for all realistic dimensions $d=1,2,3$, and leave $d\geq4$ as conjecture.

\textit{NUG phases with $\mathcal{I}_{G_\text{onsite}}\leq1$ on spin-1/2 chains.---}To show the existence of such a NUG phase, 
we consider a standard Heisenberg antiferromagnetic (HAF) spin-1/2 chain with dimerization strength $\delta\in[-1,1]$:
$\mathcal{H}_\text{HAF}^{d=1}(\delta)\equiv\sum_{j}[1+(-1)^j\delta]\vec{S}_j\cdot\vec{S}_{j+1}$,
which is gapless if $\delta=0$ while uniquely-gapped for $\delta\neq0$.
{Here we fix a sublattice (odd, even) structure so that $\delta$ and $-\delta$ are inequivalent.}
We will see that {the existence of} this ($\mathcal{I}^P_{G_\text{onsite}}=1$) gapless point on the special one-dimensional phase diagram $\delta\in[-1,1]$ actually implies $\mathcal{I}_{G_\text{onsite}}=1$ {when} we include all $G_\text{onsite}$-symmetric interaction parameters into the parameter space, shown in the following proof by contradiction.

{Let us assume that, upon including} some other $G_\text{onsite}$-symmetric interaction say $J'$,
$\mathcal{I}_{G_\text{onsite}}$ of that gapless point increases to $2$ as in FIG.~\ref{method}(b).
Then there is a {connected} path of uniquely-gapped Hamiltonians from $\mathcal{H}^{d=1}_\text{HAF}(\delta=+1)$ to $\mathcal{H}^{d=1}_\text{HAF}(\delta=-1)$ on the enlarged parameter space.
This path is an adiabatic transformation between those two Hamiltonians, {along} which the gap does not close.
However, {finding such a path is} impossible since $\mathcal{H}^{d=1}_\text{HAF}(\delta=+1)$ and $\mathcal{H}^{d=1}_\text{HAF}(\delta=-1)$ belong to distinct $G_\text{onsite}$-SPT phases classified by $\mathbb{Z}_2$~\cite{Chen-Gu-Wen_classification2010,Pollmann:2012aa} {realized by spin-1/2's~\cite{Nakamura:2002aa}}.
Indeed,
with an open boundary at the first site,
there is a single undimerized spin-1/2 for $\mathcal{H}^{d=1}_\text{HAF}(\delta=+1)$ which signals the nontrivial $G_\text{onsite}$-SPT phase, while the absence of spin-1/2 boundary state for $\mathcal{H}^{d=1}_\text{HAF}(\delta=-1)$ implies the trivial phase.
Then the presumption is false.
The above argument further implies that this gapless point keeps extending to higher-dimensional NUG phases as we include more and more parameters into the parameter space.
Any point in this extending {NUG} phase has $\mathcal{I}_{G_\text{onsite}}\leq1$ to ensure the non-existence of {any adiabatic path connecting $\mathcal{H}^{d=1}_\text{HAF}(\delta=+1)$ and $\mathcal{H}^{d=1}_\text{HAF}(\delta=-1)$},
so we call it a phase with $\mathcal{I}_{G_\text{onsite}}\leq1$ {for} short.

\textit{General construction of the topological invariant.---}
Now we generalize the above method {determining} NUG phases with $\mathcal{I}_{G_\text{onsite}}\leq1$ to {cases with} higher $\mathcal{I}_{G_\text{onsite}}$'s.
Let us assume that we can find a $(m-1)$-dimensional sphere $S^{m-1}$~\footnote{Throughout this Letter, sphere is used in the topological sense; no particular shape or location in the parameter space is specified.}
in a selected $m$-dimensional parameter space, such that each Hamiltonian on this sphere is uniquely gapped.
If the sphere is \textit{non-contractible}, that is, if it cannot be adiabatically deformed/shrunk to a point without sweeping any NUG Hamiltonian even after we enlarge {the dimensionality of the parameter space by introducing} \textit{arbitrary more} $G_\text{onsite}$-symmetric interaction parameters, 
then there must be at least one NUG Hamiltonian inside the sphere $S^{m-1}$.
Each time we include one more arbitrary interaction parameter {(i.e., one more axis in the parameter space)},
this NUG point/phase must extend to a phase of one dimension higher without termination; otherwise the sphere would be contractible in the enlarged parameter space, contradicting
the assumption.
Thus any point in this {NUG} phase must satisfy $\mathcal{I}_{G_\text{onsite}}\leq m$.
Specifically,
the earlier proof on the spin chain corresponds to $m=1$, 
making use of the non-contractible $S^{0}\cong\{\mathcal{H}^{d=1}_\text{HAF}(\delta=+1) ,\mathcal{H}^{d=1}_\text{HAF}(\delta=-1)\}$.

In order to diagnose the non-contractibility, 
we will find and assign a topological invariant $Q$ to each $S^{m-1}$ with the following property: 
if two $S^{m-1}$'s can be deformed to each other in some (maybe enlarged) parameter space without passing through any NUG Hamiltonian, 
then their $Q$'s are equal {[see the two loops with $Q_2=Q_1$ in FIG.~\ref{method}(c)]}.
Therefore, 
such a topological invariant serves as a non-contractibility {detector}; 
if a $S^{m-1}$ does not have the same $Q$ as a contractible sphere, it must be non-contractible.

Now we describe the assignment of $Q$ for $m=2$.
Namely, we consider a closed loop $S^{m-1}=S^1$ in the parameter space, along which the Hamiltonians are uniquely gapped.
We first decompose the lattice coordinate into one vertical ($V$) direction and horizontal ($H$) directions: 
$\vec{r}=(r_V,\vec{r}_H)$ with lengths $L_V$ and $L_{H;1},\cdots, L_{H;d-1}$, separately, under a periodic boundary condition (PBC).
Let us temporarily consider $d=2$, i.e., the vector $\vec{r}_H$ {is} a single number $r_H$.
Each loop of uniquely-gapped Hamiltonians can be denoted by $\mathcal{H}[\vec{J}(\tau)]$, where $\vec{J}$ is a compact notation of \textit{all} $G_\text{onsite}$-symmetric interaction parameters and $\tau$ is the loop parameter that can be freely parameterized such that {$\tau\in[0,1]$ with $\mathcal{H}[\vec{J}(\tau=0)]=\mathcal{H}[\vec{J}(\tau=1)]$}.
Then we construct a \textit{new} $r_H$-dependent {Hamiltonian}
\begin{eqnarray}
\label{loop}
\bar{\mathcal{H}}(\vec{r})\equiv\mathcal{H}[\vec{J}{(r_H/L_H)}]
\end{eqnarray}
by replacing the loop parameter $\tau$ with $r_H/L_H$.
{We expect that $\bar{\mathcal{H}}(\vec{r})$ is also uniquely gapped} because {we can choose a large-enough $L_H$ such that $|\partial_{r_H}\vec{J}|=L_H^{-1}|\partial_{\tau}\vec{J}|\ll\Delta_\tau$ and then} the gap $\Delta_\tau$ of the Hamiltonians $\mathcal{H}[\vec{J}(\tau)]$ along the loop still holds in $\bar{\mathcal{H}}(\vec{r})$ that is spatially adiabatically deformed.
We view the $r_H$-dependent $\bar{\mathcal{H}}(\vec{r})$ as a (quasi-)one-dimensional system along the vertical direction $\hat{x}_V$, which is formally an anisotropic thermodynamic limit $L_V\gg L_H\gg1$.
Since $\bar{\mathcal{H}}(\vec{r})$ also respects $G_\text{onsite}$,
it belongs to either the nontrivial one-dimensional $G_\text{onsite}$-SPT phase or the trivial one.
Let us assign $Q=-1$ to the original \textit{loop} $\mathcal{H}[\vec{J}(\tau)]$ if the corresponding $\bar{\mathcal{H}}(\vec{r})$ is in the nontrivial $G_\text{onsite}$-SPT phase while $Q=+1$ to the loop $\mathcal{H}[\vec{J}(\tau)]$ if $\bar{\mathcal{H}}(\vec{r})$ is in the trivial SPT phase.

Indeed, $Q$ constructed above qualifies as a topological invariant; if two loops $\mathcal{H}_0[\vec{J}(\tau)]$ and $\mathcal{H}_1[\vec{J}(\tau)]$ are assigned by distinct $Q_0\neq Q_1$,
they cannot be deformed continuously to each other.
{Suppose that}
we could find a series of loops $\mathcal{H}_s[\vec{J}(\tau)]$ with $s\in[0,1]$ deforming $\mathcal{H}_{0}[\vec{J}(\tau)]$ to $\mathcal{H}_{1}[\vec{J}(\tau)]$.
Each intermediate loop $\mathcal{H}_s[\vec{J}(\tau)]$ gives a $r_H$-dependent Hamiltonian $\bar{\mathcal{H}}_s(\vec{r})$ as~(\ref{loop}). 
Then $\bar{\mathcal{H}}_s(\vec{r})$ is an adiabatic path (parametrized by $s$) of uniquely-gapped Hamiltonians connecting {$\bar{\mathcal{H}}_0(\vec{r})$ and $\bar{\mathcal{H}}_1(\vec{r})$ belonging to two distinct SPT phases ($Q_0\neq Q_1$),}
which is impossible.

\begin{figure}[t]
\centering
\includegraphics[width=8.9cm,pagebox=cropbox,clip]{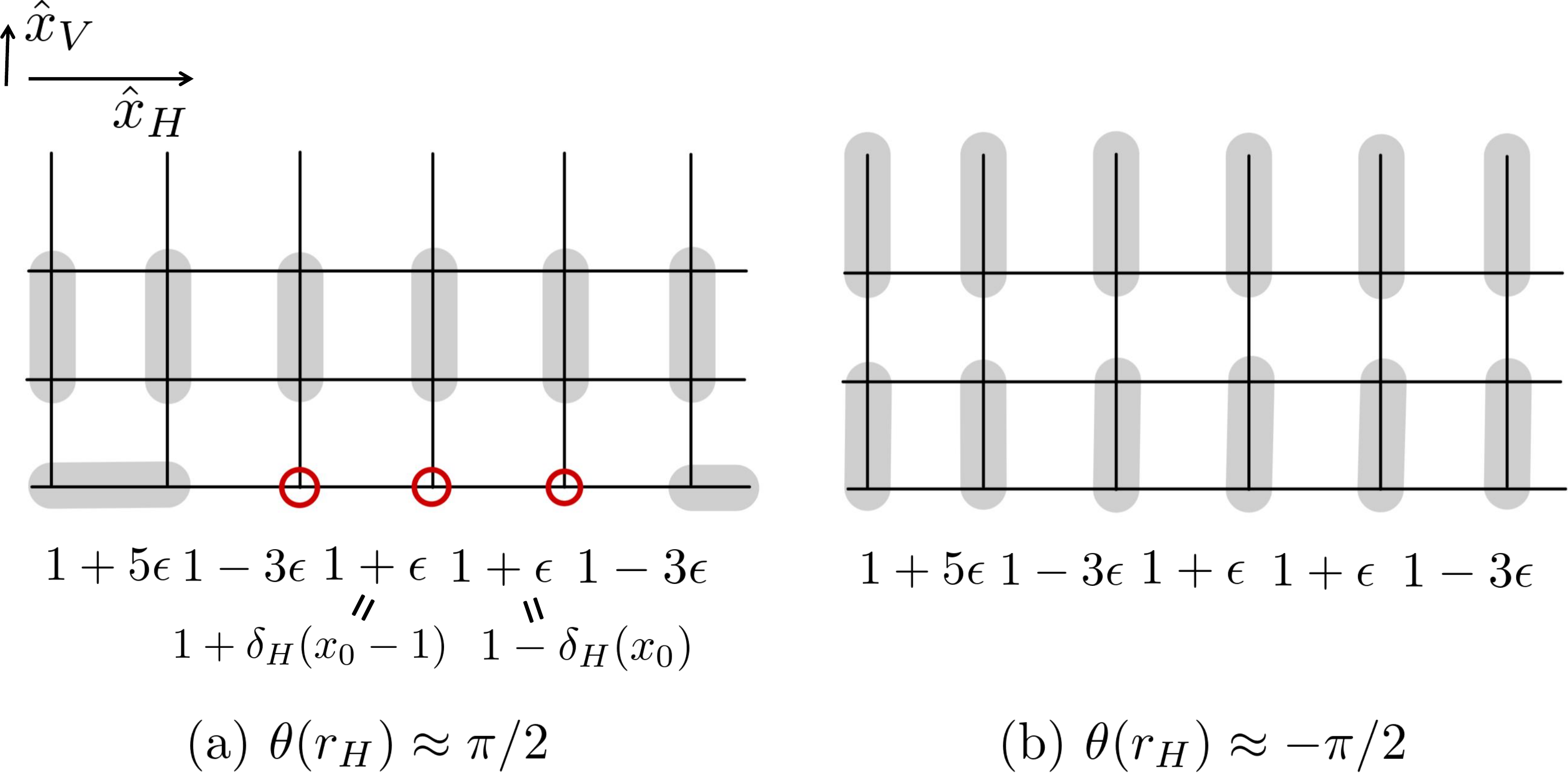}
\caption{(a) The definition of $\theta(r_H)$ and the featureless phases along the loop within which gapless points take place. (b) An odd number of undimerized spins on the boundary $r_V=1$ where $\theta(r_H)$ changes from $\pi/2-\epsilon$ to $\pi/2+\epsilon$ around $r_H=x_0$. (c) Boundary spins are dimerized into bulk where $\theta(r_H)\approx-\pi/2$.}
\label{spin_polarization}
\end{figure}

\textit{NUG phases with $\mathcal{I}_{G_\text{onsite}}\leq2$ in spin-1/2's on the square lattice.---}
Let us apply the  above general construction to the concrete system of spin-1/2's on the square lattice.
We first consider a potentially non-contractible loop in the parameter space of {the dimerized Heisenberg antiferromagnet}
\begin{eqnarray}\label{hafd=2}
\mathcal{H}_\text{HAF}^{d=2}(\delta_H,\delta_V)\!&=&\!\!\sum_{\vec{r};m=H,V}[1+(-1)^{r_m}\delta_m]\vec{S}_{\vec{r}+\hat{x}_m}\cdot\vec{S}_{\vec{r}},
\end{eqnarray}
where we {introduce the} dimerization strengths $|\delta_{H,V}|\leq1$ along both directions to span a two-dimensional parameter space {shown in} FIG.~\ref{spin_polarization}(a).
The Hamiltonians {near} the four sides $\delta_H=\pm1$ and $\delta_V=\pm1$ {of the square in the parameter space} are in decoupled ladder or decoupled four-spin-plaquette phases, which are all uniquely-gapped~\cite{Chitov:2008aa},
{while} antiferromagnetic long-range orders occur {deep inside the square}~\cite{Katoh:1993aa}.
Then we consider the loop parametrized by $\tau\in[0,1]$ winding along this square boundary for $W$ times:
$\bar{\theta}(\tau=1)=\bar{\theta}(\tau=0)+2\pi W$,
where $\bar{\theta}(\tau)\equiv\text{Arg}[\delta_H(\tau)+\sqrt{-1}\delta_V(\tau)]$ in FIG.~\ref{spin_polarization}(a).

Then we assign $Q_W=\pm1$ to this loop depending on whether its corresponding $r_H$-dependent Hamiltonian $\bar{\mathcal{H}}(\vec{r})\equiv\mathcal{H}[\theta(r_H)]$ with $\theta(r_H)\equiv\bar{\theta}(r_H/L_H)$ is $G_\text{onsite}$-SPT trivial or not.
To do so, we apply the bulk-edge correspondence with an open boundary perpendicular to the vertical direction: cutting all the bonds connecting $r_V=L_V$ and $r_V=1$ and
counting the total number of undimerized spin-1/2's along $r_V=1$ in FIG.~\ref{spin_polarization} as follows.

Such undimerized spins can {take place} where $\delta_H(r_H)$ changes sign, i.e., $\theta(r_H)=\pm\pi/2\mod2\pi$, so that they cannot be dimerized by neighboring spins along the horizontal direction.
Thus, we only need to focus around $r_H$ where $\theta(r_H)=\pm\pi/2-\epsilon$ changes to $\theta(r_H)=\pm\pi/2+\epsilon$ with $\epsilon\approx0^+$.
However,
when $\theta(r_H)\approx-\pi/2$, we have $\delta_V\approx-1$ so the boundary spins there are {readily dimerized} along the vertical direction as in FIG.~\ref{spin_polarization}(c).
On the other hand at $\theta(r_H)\approx+\pi/2$,
the undimerized spins are decoupled from the bulk spins.
We show a special paradigm of $\theta(r_H)=\pi/2+\epsilon$ changing to $\theta(r_H)=\pi/2-\epsilon$ in FIG.~\ref{spin_polarization}(b) where the boundary hosts $3$ undimerized spin-1/2's.
In general, the number of these spin-1/2's is always odd, since that sign-changing point is exactly the interface between two distinct $G_\text{onsite}$-SPT chains along the boundary{;} this interface must host an odd number of spin-1/2's, independent of lattice details~\cite{Hagiwara:1990aa,Glarum:1991aa}.
Thus
the total number of undimerized boundary spin-1/2's is the winding number $W \mod2$, which means
$Q_W=\exp(i\pi W)=\exp\left(i\pi\int_{r_H}{d\theta(r_H)}/{2\pi}\right)$,
because the bulk-edge correspondence of $G_\text{onsite}$-SPT {phases} implies that the nontrivial (trivial) phase hosts an odd (even) number of spin-1/2's on the boundary~\footnote{More strictly speaking, a nontrivial (trivial) one-dimensional $G_\text{onsite}$-SPT phase hosts a (non-)projective representation at the boundary.}.
It means that the loop with $W=1$ is non-contractible, thereby detecting a NUG phase with $\mathcal{I}_{G_\text{onsite}}\leq2$
extended without termination from a gapless point in FIG.~\ref{spin_polarization}(a) as more interaction parameters are included.
{The loop is similar to a Floquet system~\cite{Else:2016aa,Keyserlingk:2016aa,Potter:2016aa,Roy:2016aa,Roy:2017aa,Shiozaki:2021aa} but spatially periodic here.
The following multi-variable extension does not have Floquet analogs.}

\textit{NUG phases with $\mathcal{I}_{G_\text{onsite}}\leq3$ on spin-1/2 cubic lattices.---}
Our construction of the bound $\mathcal{I}_{G_\text{onsite}}\leq m$ based on a topological invariant on $S^{m-1}$ in the parameter space can be extended to $m>2$.
Let us illustrate the construction for $m=3$, with respect to the spin-1/2's on the cubic lattice.
We consider the following typical Hamiltonian: $(J_V,J_{H;1,2}>0)$
\begin{eqnarray}\label{hafd=3}
&&H_\text{HAF}^{d=3}(\vec{\delta}_H,\delta_V)=\sum_{\vec{r}}\biggl\{
J_V[1+(-1)^{r_V}\delta_V]\vec{S}_{\vec{r}+\hat{x}_V}\cdot\vec{S}_{\vec{r}}
\nonumber\\
&&\qquad\qquad\qquad+\sum_{n=1}^{2}J_{H;n}[1+(-1)^{r_{H;n}}\vec{\delta}_{H;n}]\vec{S}_{\vec{r}+\hat{x}_{H;n}}\cdot\vec{S}_{\vec{r}} \biggr\}\!,\nonumber
\end{eqnarray}
where, in addition to dimerization strengths along each direction, the antiferromagnetic {exchange couplings $J_V$ and $J_{H;1,2}$ are also necessary control parameters} in the following construction of a non-contractible sphere.

We start from the {cube surface in the parameter space, topologically} a sphere, consisting of six faces {defined by one of the following conditions:} $\delta_V=\pm1,\,\delta_{H;1}=\pm1,\,\delta_{H;2}=\pm1$.
The Hamiltonians on twelve edges belong to the phase of decoupled four-leg spin tube, which are uniquely-gapped~\cite{Fuji:2016aa}.
Thus, we simply set $J_{V}=J_{H;1,2}=1$ {on the twelve edges.
By contrast,}
as we approach each face center {where the Hamiltonian is reduced to a set of decoupled bilayer antiferromagnets},
antiferromagnetic long-range order (LRO) emerges if we keep $J_{V,H;1,2}=1$~\cite{Matsuda:1990aa}.
{To avoid LRO,
we gradually increase $J_V$ from 1 to, say, 3 on the faces $\delta_V=\pm1$ when approaching from an edge to the face center so that the Hamiltonians on this entire face are} uniquely-gapped.
This is possible because no phase transition occurs even when $J_V$ is increased from $1$ to infinity {near the edges}~\cite{Fuji:2016aa}.
Similarly,
we {increase $J_{H;j}$ from edges to the center} on each face $\delta_{H;j}=\pm1$.
The resultant parameter surface, denoted as $\mathcal{C}$, of this parameter cube~\footnote{The parameter setting in the interior of the cube can be done rather arbitrarily.} is {topologically a sphere $S^2$ to be shown non-contractible below}.

\begin{figure}[t]
\centering
\includegraphics[width=8.7cm,pagebox=cropbox,clip]{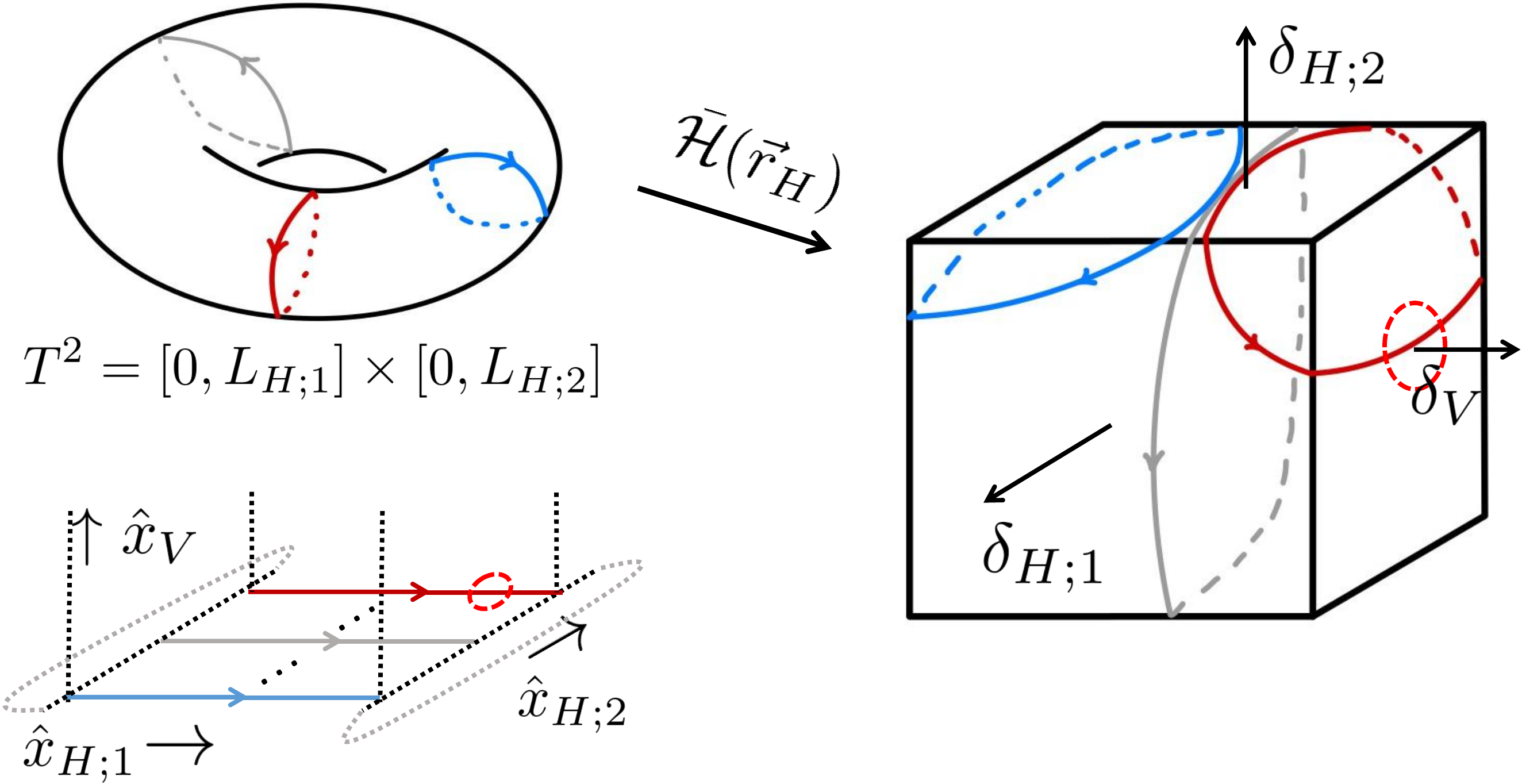}
\caption{$\bar{\mathcal{H}}(r_{H;1},r_{H;2})$ wraps the torus $T^2$ around $\mathcal{C}\cong S^2$. {The position of undimerized boundary spins is indicated by the circle} ``{\color{red}$\circ$}''.}
\label{wrapping_number}
\end{figure}

Following the general construction,
we assign $Q$ to this sphere through a $\vec{r}_H$-dependent Hamiltonian $\bar{\mathcal{H}}(\vec{r})=\bar{\mathcal{H}}(r_{H;1},r_{H;2})$ as follows.
Since we have PBC along the horizontal directions,
$\bar{\mathcal{H}}(r_{H;1},r_{H;2})$ can be seen as a mapping from the torus $T^2=[0,L_{H;1}]\times[0,L_{H;2}]$ to the parameter-cube surface $\mathcal{C}\cong S^2$ constructed above.
Then we take $\bar{\mathcal{H}}(r_{H;1},r_{H;2})$ as a typical mapping which wraps once the ``sphere'' $\mathcal{C}$ by $T^2$ as in FIG.~\ref{wrapping_number}.
Next, we determine its $Q$ by studying which one-dimensional $G_\text{onsite}$-SPT phase $\bar{\mathcal{H}}(\vec{r})$ belongs to when viewed as a one-dimensional system along {the vertical unit vector $\hat{x}_V$}.
Again,
we make an open boundary at $r_V=1$ perpendicular to the vertical direction $\hat{x}_V$ and count undimerized spins there.
{Each loop in FIG.~\ref{wrapping_number} is a two-dimensional analog of} the boundary in FIG.~\ref{spin_polarization}. {A similar consideration gives that}
undimerized spins can only take place on each loop where {both $\delta_{H;1}$ and $\delta_{H;2}$ change sign and} $\delta_V>0$.
Such a situation occurs exactly once for $\bar{\mathcal{H}}(r_{H;1},r_{H;2})$ circled in FIG.~\ref{wrapping_number}.
Thus odd number(s) of undimerized spins are dangling at the certain position on the horizontal boundary, which implies that $Q=-1$.
In general, we can {wrap} the torus $T^2$ around the cube $\mathcal{C}$ repeatedly for $\mathcal{W}$ times by $\bar{\mathcal{H}}^{(\mathcal{W})}(r_{H;1},r_{H;2})$, and thus the topological-invariant assignment for this general wrapping, {a higher-dimensional analog of $Q_W$,} is
$Q_{\mathcal{W}}=\exp(i\pi\mathcal{W})=\exp\left(i\pi\int_{\vec{r}_H}{d\Omega_2(\vec{r}_H)}/{4\pi}\right)$,
where $d\Omega_2$ is the differential solid angle of the ``sphere'' $\mathcal{C}$ spanned by the vector $(\delta_{H;1},\delta_{H;2},\delta_V)$.
Specifically,
it means that the cube surface $\mathcal{C}$, which is of $\mathcal{W}=1$, cannot be contracted continuously to a point ($\mathcal{W}=0$) due to their different $Q_\mathcal{W}$'s.
Thus its non-contractibility signals an extending NUG phase with $\mathcal{I}_{G_\text{onsite}}\leq3$, as announced.

\textit{NUG phases $\mathcal{I}_G\leq d-k$ with $k$ translations.---}
So far we have shown the ingappability theorem in realistic dimensions $d=1,2,3$ without translations imposed~\footnote{For example, with any coupling that respects $G_\text{onsite}$ but breaks translation symmetry, the inequality $\mathcal{I}_{G_\text{onsite}}\leq d-k$ of the NUG phase still holds.}.
{When $G=G_\text{onsite}\times\mathbb{Z}^k$ is imposed on the $d$-dimensional Hamiltonian with $\mathbb{Z}^k$ translations along directions~($k<d$),
we can construct a non-contractible $S^{d-k-1}$ in the parameter space by stacking identical $(d-k)$-dimensional $\bar{\mathcal{H}}(\vec{r})$
along those $k$ directions.
(When $d=k+1$,
it is the weak $G$ SPT~\cite{Fu:2007aa}.)
It proves our statement of the existence of NUG phases with $\mathcal{I}_G\leq d-k$.}

\textit{Critical point and symmetry-protected gapless phases.---}
When the NUG phase is critical, it is described in terms of a Renormalization-Group (RG) fixed point or a scale-invariant field theory.
The gappability index $\mathcal{I}_G$ strongly restricts its nature:
\textit{A critical point in the NUG phase with a finite $\mathcal{I}_G$ can have at most $\mathcal{I}_G$ (marginally) relevant operators gapping out the system to
yield a unique ground state.}

In particular, {when} $d=1$ and $k=0$ with $G=\text{SO}(3)$,
SU$(2)$ level-$1$ Wess-Zumino-Witten model describing this critical point indeed possesses only $\mathcal{I}_G =d-k = 1$
relevant operator~\footnote{In terms of the SU(2) level-$1$ Wess-Zumino-Witten model, the local relevant perturbation is of the form of $\text{tr}(g)$ with $g$ a SU$(2)$ matrix-valued field.
The next leading SO(3) invariant operator is $\mathrm{tr}g^2$, which can be marginally relevant. However, when it induces a gap, the ground states are doubly degenerate.}.
{The low-energy effective field theory for HAF in $d>1$ should have $\mathcal{I}_{G}\leq d$.
This gives a strong constraint on the possible candidates for such an effective field theory.}
This observation should be extended to systems only with discrete symmetries~\cite{Fuji:2016aa}.
On the other hand, there are many open questions about field theories describing critical points in higher dimensions.
The present result gives powerful constraints on the possible field theory, and an insight into the role of translational symmetries on {the critical-phase stability}.

Additionally,
$\mathcal{I}_G$ can classify gapless critical phases, {which are}
inaccessible by {classification theory of conventional gapped topological phases} due to the absence of gaps.
{Gapless critical phases} are protected by symmetry $G$ in that a lower symmetry generically results in a larger $\mathcal{I}_G$,
and ``trivial'' phases correspond to $\mathcal{I}_G=\infty$.
Moreover,
$Q_{W,\mathcal{W}}$'s do not exclude {$W,\mathcal{W}\in\mathbb{Z}_\text{even}$ doubly winding/wrapping around the loop/cube} to be contractible.
Thus this even number and $\mathcal{I}_G$ form a {finer} symmetry-protected classification of the gapless phases generalizing previous proposals by LSM ingappabilities~\cite{Furuya:2017aa,Yao:2019aa,Yao:2020PRX}.


We thank X.-G. Zhou for helpful discussions. This work was supported in part by MEXT/JSPS KAKENHI Grant Nos.\ JP17H06462, JP19H01808 and JP19K03680, and JST CREST Grant No.\ JPMJCR19T2.

%
%
%


%

\end{document}